\documentclass[osajnl,twocolumn,showpacs,superscriptaddress,10pt]{revtex4-1}
\usepackage{amsmath,amssymb,graphicx}
\usepackage{dcolumn}    
\usepackage{bm}         
\usepackage[caption=false]{subfig}
\begin{document}

\title{Hyperbolic metamaterial as a tunable near-field spatial filter for the implementation of the active plasmon injection loss compensation scheme}

\author{Anindya Ghoshroy}
\author{Xu Zhang}
\author{Wyatt Adams}
\author{Durdu \"O. G\"uney}\email{Corresponding author: dguney@mtu.edu}
\affiliation{Department of Electrical and Computer Engineering , Michigan Technological University, 1400 Townsend Dr, Houghton, MI 49931-1295, USA}

\begin{abstract}
We present how to physically realize the auxiliary source described in the recently introduced active plasmon injection loss compensation scheme for enhanced near-field superlensing. Particularly, we show that the characteristics of the auxiliary source described in the active plasmon injection scheme including tunable narrow-band and selective amplification via convolution can be realized by using a hyperbolic metamaterial functioning as a near-field spatial filter. Besides loss compensation, the proposed near-field spatial filter can be useful for real-time high resolution edge detection.

\end{abstract}
\maketitle

\section{Introduction}

Our ability to control electromagnetic fields with metamaterials has flourished since the turn of the century and has in turn engendered a myriad of previously unthought of applications. As opposed to the electromagnetic properties of naturally occurring materials, the properties of metamaterials stem primarily from their subwavelength structural details rather than their chemical properties alone. By carefully controlling these subwavelength features, one can fabricate an artificial material with electromagnetic properties which are very rare and sometimes impossible to find in nature. One important application of metamaterials is in the field of near-field superlensing. The near-field optics coupled with plasmonics and metamaterials has a wide range of implications from subdiffraction imaging \cite{adams2016review} to enhanced absorption \cite{rockstuhl2008absorption,gwamuri2016new,ahmadivand2015enhancement}. In the context of imaging, the near-field contains information about the subwavelength features of an object and is evanescent in nature. Pendry envisioned \cite{pendry2000negative} that a slab of negative index material (NIM) can be used to amplify these evanescent waves and renewed interest in the obscure idea of NIMs  first conceived by Veselago \cite{veselago1968electrodynamics} in the late 1960s.

In the years that followed, a NIM was realized \cite{shelby2001experimental} for the first time by Shelby, et. al followed by the demonstration of a near-field superlens that exhibits imaging beyond the diffraction limit \cite{fang2003imaging,melville2005super}. Subsequently, it was realized that the presence of inherent material losses substantially degrade the performance of superlenses \cite{Podolskiy:05,smith2003limitations,PhysRevLett.95.137404}. A robust loss compensation scheme was clearly necessary and in the ideal case should be completely independent of the object. Efforts to overcome this problem led to the development of new approaches at loss compensation which employed gain medium \cite{PhysRevLett.105.127401,xiao2010loss,Nezhad:04}, non-linear effects \cite{Popov:06}, and geometric tailoring \cite{guney2009reducing}. However, these approaches introduced additional complexities. For example, using gain medium makes it difficult to preserve the amplitude and phase relationships between the fields in time and space which is crucial especially for imaging applications.

We have been attempting to develop a new loss compensation scheme where the goal is to use an external \emph{``auxiliary"} illumination to compensate losses in the material. This was initially conceptualized in \cite{sadatgol2015plasmon}, where the losses suffered by a normally incident wave were compensated by a coherent superposition with an auxiliary field. Although the method was studied in detail for a single wavevector, it inspired two important questions. If one could develop  a similar technique, where an auxiliary source provides compensation for a large band of wavevectors, would it be possible to perfectly reconstruct the original object in an imaging scenario without having any prior knowledge of the object? If so, could the technique be applied to different near-field imaging systems such as those employing NIMs or plasmonic lenses using, for example, silver \cite{fang2005sub} and silicon carbide (SiC) \cite{taubner2006near}, or hyperlenses \cite{jacob2006optical} under both coherent and incoherent illumination? In \cite{sadatgol2015plasmon}, we used the name ``plasmon injection ($\Pi$) scheme," referring to the above form of loss compensation which employs an external auxiliary source to amplify the decayed Fourier components propagating inside the lossy plasmonic metamaterial. It has been envisioned that the amplitude of each Fourier component to be provided by the auxiliary could be estimated from the transfer function of the imaging system. Subsequent efforts were directed at answering the second question. It was demonstrated theoretically that the technique in general could be applied to different imaging systems to improve their resolution limits \cite{adams2016bringing,zhang2016enhancing,Zhang:17,Adams:17}. However, no physical auxiliary source was considered. It was simply assumed that one already has the means to amplify an arbitrary Fourier component as proposed in \cite{sadatgol2015plasmon}. Before attempting to physically realize the auxiliary source, it was important to understand what properties the auxiliary should possess in order to compensate losses in a realistic noisy imaging system. Continuing efforts \cite{Ghoshroy:17} showed that the auxiliary must provide ``selective amplification" to a narrow-band of high-frequency Fourier components to avoid large noise amplification. Additionally, to recover Fourier components of the object buried in the noise, the auxiliary source has to be constructed by the physical convolution of the object field.

In this work, we show that metal-dielectric systems with a hyperbolic dispersion operating as a tunable spatial filter can be used to construct the auxiliary source and preserve the necessary characteristics shown in \cite{Ghoshroy:17}. Selective amplification property relies on the selective spatial filter functionality of such physical systems. Since the auxiliary is to be applied in the deep subwavelength region in the reciprocal space, the proposed spatial filter is designed to strongly suppress the propagating modes while allowing the transmission of a tunable band of evanescent modes. Layered metal-dielectric systems with hyperbolic dispersion are one possible solution for such spatial filters since they are known to support wavevectors with large transversal components exceeding the diffraction limit. This is due to the presence of coupled surface plasmon polariton (SPP) modes at the interfaces. By modifying the permittivities of the constituent materials one can control the eigenmodes supported by the system.

\section{Theory}

In \cite{Ghoshroy:17}, the auxiliary source is defined as the convolution of the object field with a function whose Fourier transform, $P(k_y)= \mathcal{F}\{P(y)\}$, is a Gaussian. In the reciprocal space, the product of the object spectrum with this Gaussian represents the amplification provided to a band of Fourier components. The physical realization of this product requires the following properties. First, one needs to find a material whose transmission spectrum for different spatial frequencies has a functional form  with properties similar to $P(k_y)$ i.e., a relatively high transmission peak at a tunable centre frequency and near-zero transmission on either side especially in the lower spatial frequencies. The material should therefore possess the characteristics of a band-pass filter for high spatial frequencies in the Fourier domain. Second, the response of the material should have shift invariance in the plane perpendicular to the optical axis. A system is said to have shift invariance along a plane if its response to a point source excitation changes only in spatial position on that plane but not in functional form, as the point source traverses the excitation plane.  This shift-invariance property is necessary to relate the fields in the input and output planes with a convolution. These properties may be difficult, if not impossible, to realize with isotropic materials, but it is well known that there exists a class of metamaterials with hyperbolic dispersion which supports the propagation of evanescent modes.

Spatial filtering using hyperbolic metamaterials (HMMs) has only been loosely studied before \cite{schurig2003spatial, rizza2012terahertz} with no explicit considerations of near-field or loss compensation. Here we briefly look at some of the properties of HMMs which are relevant to this work.  A more comprehensive analysis on the properties of hyperbolic materials can be found in the following works \cite{poddubny2013hyperbolic,Shekhar2014,drachev2013hyperbolic} and the references cited within.

\begin{figure}[htbp]
\centering
\includegraphics[height=0.3\textwidth]{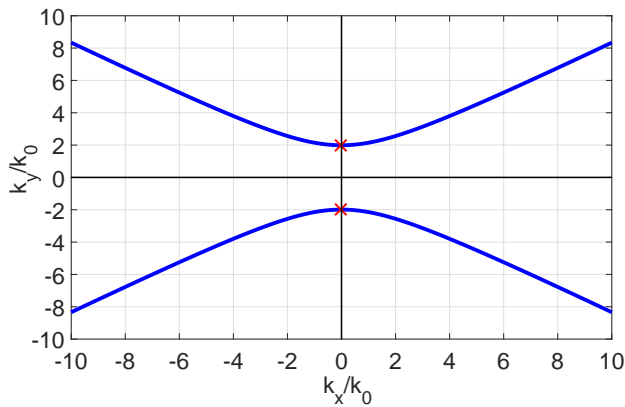}
\caption{Isofrequency contour plot of an anisotropic material with $\epsilon _{y} < 0$ and  $\epsilon _{x} > 0$ at $\lambda = 365 \ nm$.  The red markers indicate the cut-offs for the transversal wavevector components. The artificial material corresponding to the plot is realized by stacking 8 periods of alternating aluminium and quartz layers. The principal relative permittivity components are calculated with the effective medium approximation (see text for details).}
\label{fig:Fig_Hyperbolic}
\end{figure}

Consider a plane wave propagating in the xy-plane through a uniaxial crystal. The wavevector component $k_y$ represents a spatial variation along the transversal direction and the wavevector component $k_z=0$. Therefore, by looking at the sign of $k_x^2$ one can distinguish between the propagating and evanescent modes. Without loss of generality, we assume that the material is magnetically isotropic. The elements of the relative permittivity tensor in the principal coordinate system are given by
\begin{equation}
\epsilon =
  \begin{bmatrix}
    \epsilon _{x}  &     0        &     0 \\
        0        & \epsilon _{y}  &     0 \\
        0        &     0        & \epsilon _{z}
  \end{bmatrix}
\label{eq:Epsilon_tensor}
\end{equation}
with $ \epsilon _{y} = \epsilon _{z}$. The dispersion relation for the extraordinary waves can be easily determined from the eigenvalue equation and is
\begin{equation}
\frac{k_x ^2}{\epsilon _{y}} + \frac{k_y ^2}{\epsilon _{x}} = \frac{\omega ^2}{c^2 },
\label{eq:Dispersion_extraordinary}
\end{equation}
where $\omega$ is the angular frequency and $c$ is the speed of light in vacuum. Eq. \ref{eq:Dispersion_extraordinary} describes an elliptical dispersion for conventional uniaxial media. However, if the signs of the principal relative permittivity components are not the same i.e $\epsilon _{y} < 0$ and  $\epsilon _{x} > 0$, then Eq. \ref{eq:Dispersion_extraordinary} transforms into a hyperbola. The isofrequency contour of such a medium for transverse magnetic (TM) polarized light is shown in figure \ref{fig:Fig_Hyperbolic}. The choice of the parameters and the operating wavelength in the figure will be detailed later on. We should also note that the contributions from the imaginary parts of the calculated relative permittivity to the isofrequency contour in figure \ref{fig:Fig_Hyperbolic} are negligible.  From this isofrequency contour, one immediately conclude that the open form of the hyperbola allows for the propagation of modes with infinitely large transversal wavevector components, which generally leads to evanescent waves in conventional isotropic and uniaxially anisotropic media. Additionally, we make a note of the intercepts on the ordinate axis, which are crossed in figure \ref{fig:Fig_Hyperbolic}. This shows that the medium only supports the propagation of high transversal wavevector components $k_y$ beyond a certain cut-off.
An artificial metamaterial which has the above properties represents the ultra-anisotropic limit in uniaxial crystals. A negative value of relative permittivity means that the polarization response of electrons will be opposite to the direction of the incident electric field and is common in metallic elements below the plasma frequency. Therefore, in the hyperbolic regime, the elements of the relative permittivity tensor tell that the material should behave as a metal in one direction and as a dielectric in the other.

\begin{figure}[htbp]
\centering
\includegraphics[height=0.25\textwidth]{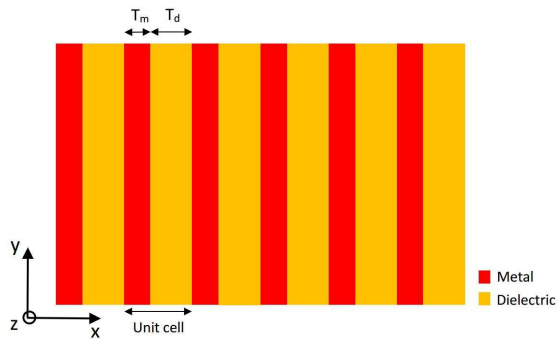}
\caption{Schematic of a metamaterial with hyperbolic dispersion (not to scale). The red and orange layers are the metallic and dielectric layers with thicknesses $T_m$ and $T_d$, respectively. The optical parameters of such a composite medium can be calculated from the effective medium approximation under the assumption that the longitudinal wavevector component $k_x << (T_m + T_d)^{-1}$.}
\label{fig:Fig_HMM_Schematic}
\end{figure}

A common type of HMMs are physically constructed by alternately stacking metallic and dielectric layers as shown in figure \ref{fig:Fig_HMM_Schematic}. Assuming that the electromagnetic parameters of individual layers are homogeneous and isotropic and the unit cell thickness is sufficiently small compared to the wavelength of the incident radiation such that the Maxwell-Garnett effective medium approximation is valid, then the system can be described as an anisotropic medium whose principal permittivities are
\begin{equation}
\frac{1}{\epsilon _{x}} = \frac{1}{1+\eta} \bigg [ \frac{1}{\epsilon _{d}} + \frac{\eta}{\epsilon _m} \bigg ]
\label{eq:Tensor_x}
\end{equation}
\begin{equation}
\epsilon _{y} = \epsilon _{z} = \frac{\epsilon _{d} + \eta \epsilon _{m}}{1+\eta},
\label{eq:Tensor_yz}
\end{equation}
where $\eta$ is the filling fraction, which is defined as the ratio of the thicknesses of the two layers
\begin{equation}
\eta = \frac{T_m }{T_d},
\label{eq:Filling_ratio}
\end{equation}
and $\epsilon _{d}$ and $\epsilon _m$ are the permittivities of the dielectric and metallic layers, respectively. The cut-off transversal wavevector components marked in the isofrequency contour in figure \ref{fig:Fig_Hyperbolic} can be expressed in terms of the effective parameters as
\begin{equation}
\frac{k_{y}^{c}}{k_o} = \pm \sqrt{\frac{(1+\eta)\epsilon _m \epsilon _d}{\epsilon _m + \eta \epsilon _d}}.
\label{eq:Cutoff_frequency}
\end{equation}
This tells that the cut-off can be tuned by changing the material and geometric parameters of the system. In fact, later on we will use Eq. \ref{eq:Cutoff_frequency} to select the available materials when constructing a spatial filter with a desired cut-off frequency.

\section{Results and Discussion}

For the rest of this paper we will use aluminium as the metallic layer and set the operating wavelength at $\lambda = 365 \ nm$. This wavelength is selected because there exist near-field imaging systems and physical sources centered at $365 \ nm$ \cite{fang2005sub}. The relative permittivity of aluminum is described with the Drude-Lorentz model \cite{rakic1998optical} and is given as
\begin{equation}
\epsilon _m(\omega) = 1 - \frac{f_o \omega _p ^2}{\omega(\omega - i\Gamma _o)} + \sum \limits ^{k} _{j=1} \frac{f_j \omega _p ^2}{(\omega _j ^2 - \omega _p ^2) - i\omega \Gamma _j },
\label{eq:Drude_Lorentz}
\end{equation}
where $\omega _p$ is the plasma frequency, $f_o$ is the oscillator strength, and $\Gamma _o$ is the damping constant, considering the free electron or intraband effects. The interband or bound-electron effects are accounted by a semiquantum model with $j$ oscillators where $\omega _j$ is the $j^{th}$ oscillator frequency, $f_j$ is the $j^{th}$ oscillator strength and $\Gamma _j$ is the $j^{th}$ damping constant. The relative permittivity of aluminum calculated from Eq. \ref{eq:Drude_Lorentz} at $\lambda = 365 \ nm$ is $\epsilon _m = -18.179 -i3.2075$. The dielectric layer is selected to be quartz $(SiO_2)$ which has a relative permittivity of $\epsilon _d = 2.2147$ at the selected wavelength \cite{gao2012exploitation}. Using these parameters and assuming that Eqs. \ref{eq:Tensor_x} and \ref{eq:Tensor_yz} are valid, the relative permittivity tensor elements of the effective anisotropic material are $\epsilon _{x} = 3.5302 - 0.0391i$ and $\epsilon _{y} = -4.5832 - 1.0692i$. In figure \ref{fig:Fig_Hyperbolic}, we used these parameters to plot the isofrequency contour for the extraordinary waves described by Eq. \ref{eq:Dispersion_extraordinary}.

\begin{figure}[htbp]
\centering
\includegraphics[height=0.5\textwidth]{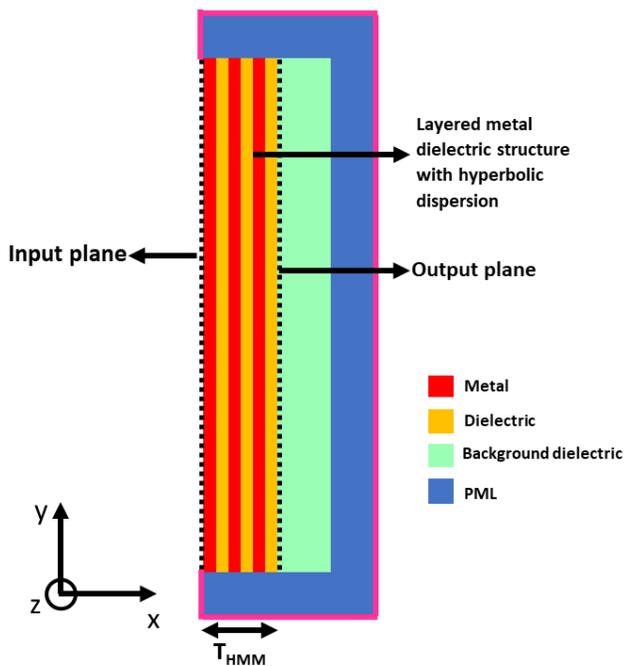}
\caption{The geometry constructed in COMSOL to perform numerical simulations (not to scale). Magnetic field polarized along the z-axis (pointing out of plane) is incident on the HMM. The response of the system is extracted on the output plane on the opposite surface of the HMM with thickness $T_{HMM}$. The red, orange, green, and blue regions are the metallic and dielectric layers of the HMM, background dielectric, and PMLs, respectively. Scattering boundary conditions are applied to the edges of the computational domain that are highlighted in pink.}
\label{fig:Fig_Simulation_geometry}
\end{figure}

To determine the transmission properties of the HMM to be employed as a near-field spatial filter for different wavevectors, we use the finite element method based commercial software package COMSOL Multiphysics. The simulation geometry on the xy-plane is shown in figure \ref{fig:Fig_Simulation_geometry}. $T_{HMM}$ is the thickness of the HMM and is set to $365 \ nm$. Considering the limits of current fabrication technologies and the validity of the effective medium approximation, we have used $8$ unit cells with the filling fraction $\eta=0.5$. Therefore, each unit cell is approximately $45 \ nm$ thick and the metallic and dielectric layers are approximately $15 \ nm$ and $30 \ nm$ thick, respectively. The HMM is assumed to be embedded in dielectric with relative permittivity $\epsilon _d = 2.5$ (see figure \ref{fig:Fig_Simulation_geometry}) which is close to the relative permittivity of the dielectric used for an experimental silver lens in \cite{fang2005sub}. The edges of the geometry are padded with perfectly matched layers (PMLs) shown in blue in figure \ref{fig:Fig_Simulation_geometry} and backed by scattering boundary conditions at the edges of the computational domain shown by the pink lines.

Since the unit cells are stacked along the x-axis, the geometry should have shift invariance in the yz-plane. Thus, we can describe the response of the system to an arbitrary field distribution using its response to the dipolar point source, which is the point spread function (PSF) of the system or the transfer function in the Fourier domain. A point source can be approximated by a Gaussian field distribution as long as the full width at half maximum (FWHM) is extremely small compared with the operating wavelength. A TM polarized Gaussian field distribution with $FWHM=5.73 \ nm$ is applied to the input plane of the HMM as shown in figure \ref{fig:Fig_Simulation_geometry} to determine the transfer function.
It is important to maintain a fairly high degree of accuracy in the calculations. Therefore, the spatial extent of the geometry along the y-axis is set to $80$ times the wavelength. This is necessary  for the shift invariance and capturing the sufficiently large extent of the field from the output plane. The data should not be abruptly truncated since this will introduce errors in the Fourier transform calculations. Additionally, due to the excitation of SPPs with large transversal wavevectors, there will be rapid field oscillations on the output plane, as well as at each metal-dielectric interface. To capture this field accurately we used $9500$ mesh elements at each interface parallel to the y-axis with the smallest mesh element being approximately equal to $3 \ nm$.

\begin{figure}[htbp]
\subfloat[\label{subfig-1:Subfig_PSF_amp}]{\includegraphics[height=0.25\textwidth]{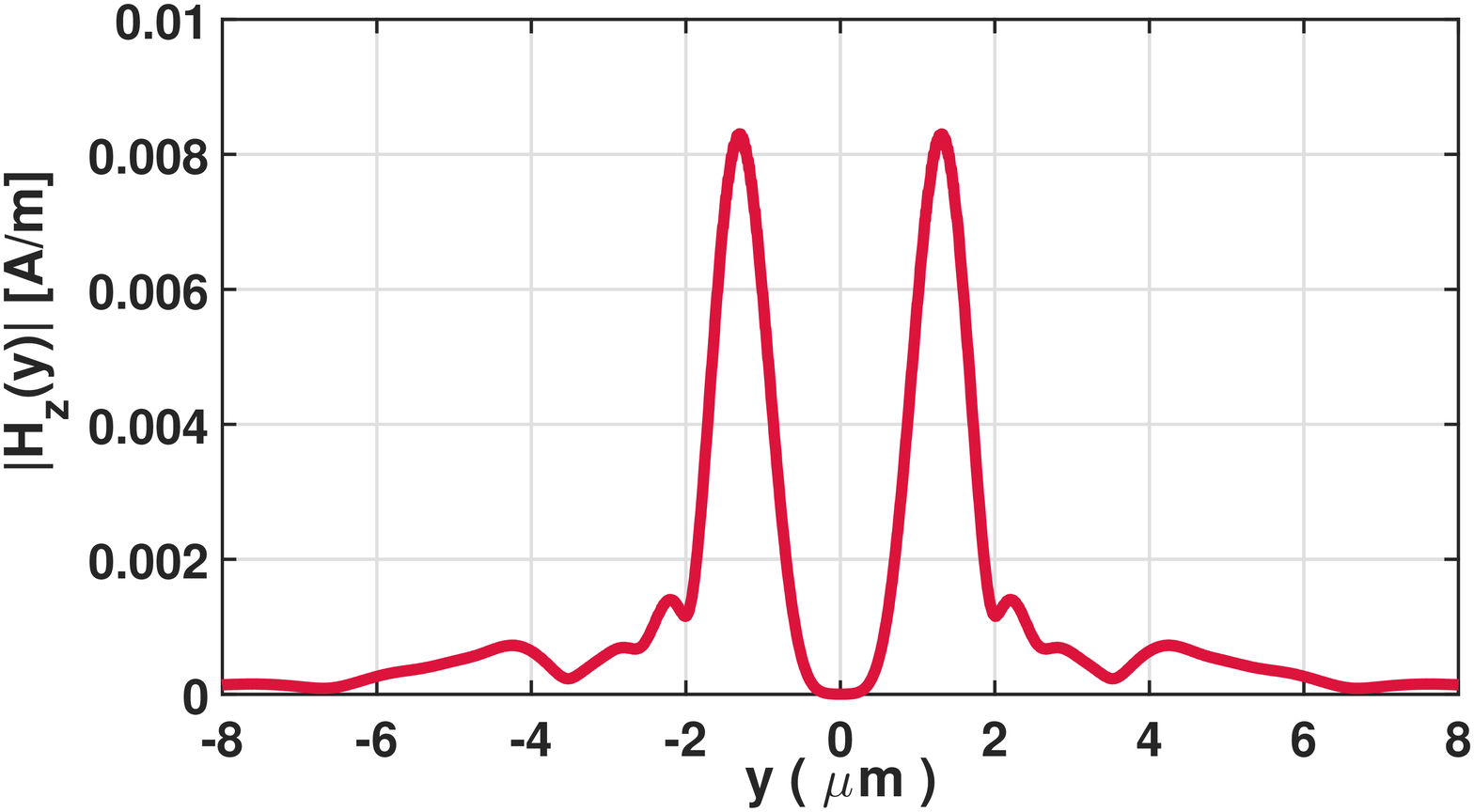}}  \\
\subfloat[\label{subfig-2:Subfig_PSF_phase}]{\includegraphics[height=0.25\textwidth]{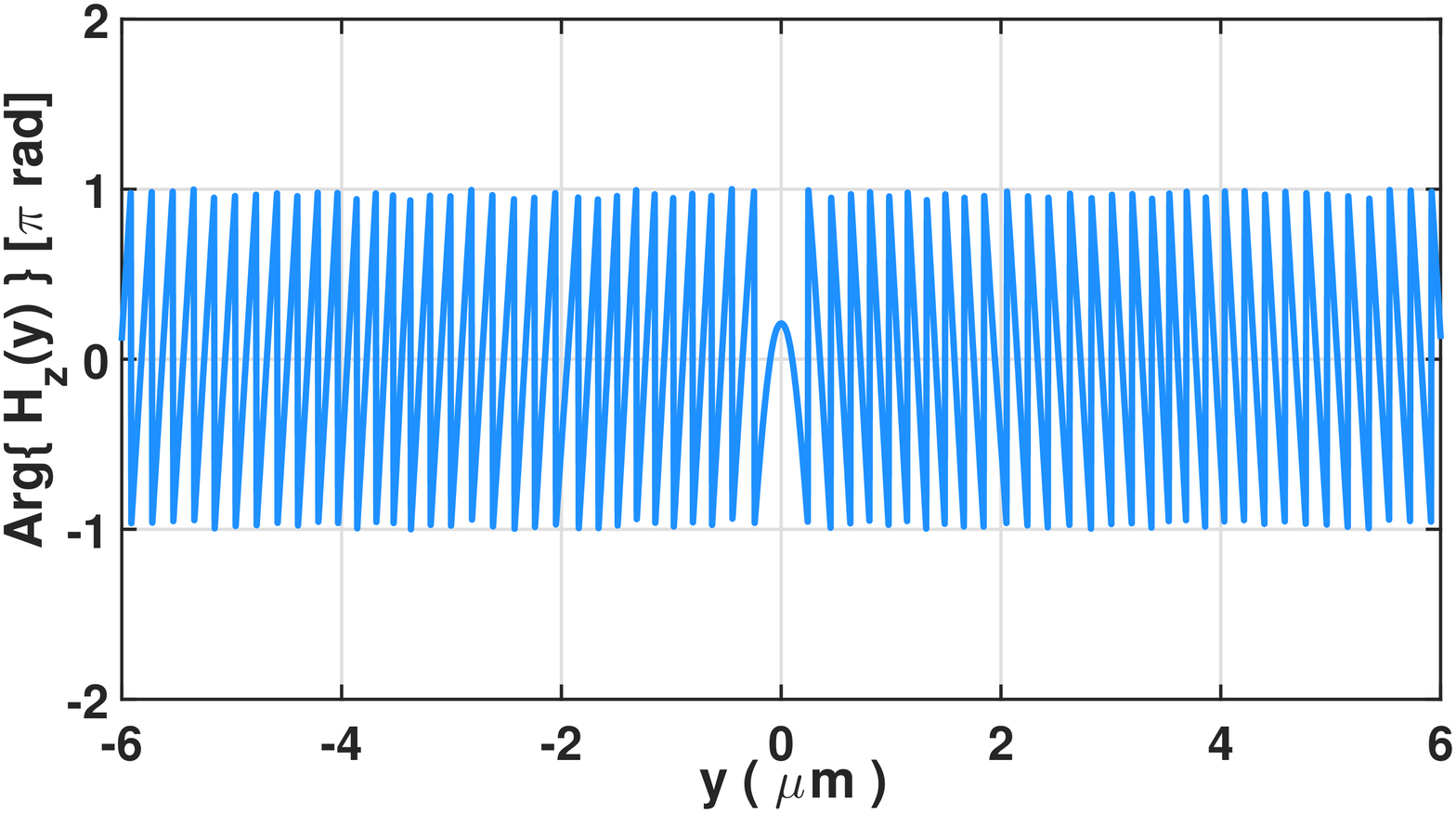}} \\
\caption{Plots of the (a) amplitude and (b) phase of the complex magnetic field distribution in response to a point source excitation. This is the PSF of the system verified by shift invariance and can be used to determine the response to any arbitrary incident excitation by using convolution.}
\label{fig:Fig_PSF}
\end{figure}

Having explained the computational subtleties in the transfer function calculation, we next analyze the response of the system to a point source excitation. The response of the multi-layered HMM structure is extracted from the output plane as a complex magnetic field distribution. The amplitude and phase of the magnetic field are plotted in figures \ref{subfig-1:Subfig_PSF_amp} and \ref{subfig-2:Subfig_PSF_phase}, respectively. If the shift invariance of the system is verified, this field distribution becomes the PSF of the system and the response to any arbitrary field distribution can be calculated from the PSF by using convolution.

\begin{figure}[htbp]
     \subfloat[\label{subfig-1:Subfig_Arbitrary_field}] {\includegraphics[height=0.25\textwidth]{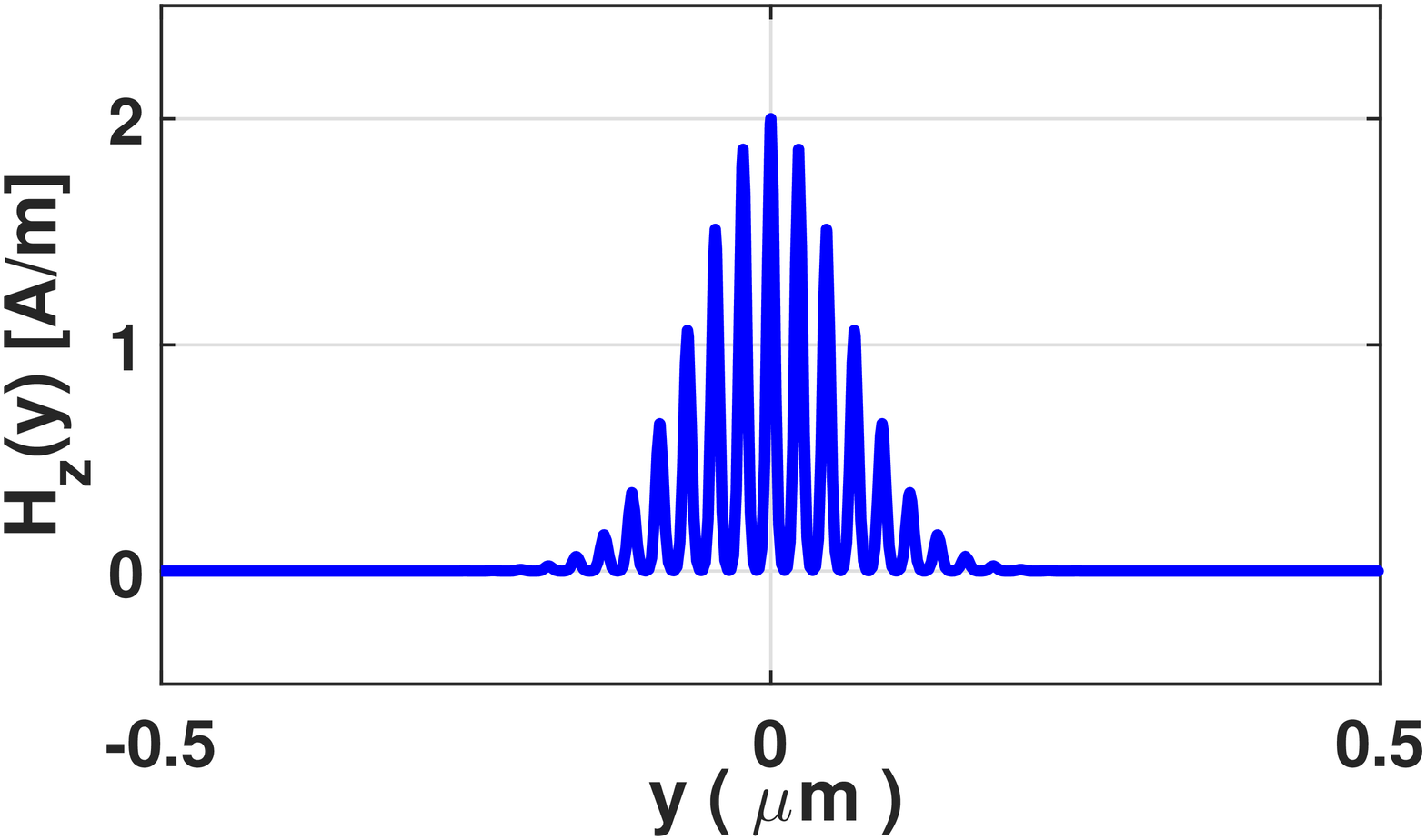}}  \\
     \subfloat[\label{subfig-2:Subfig_Convolution_real}]{\includegraphics[height=0.25\textwidth]{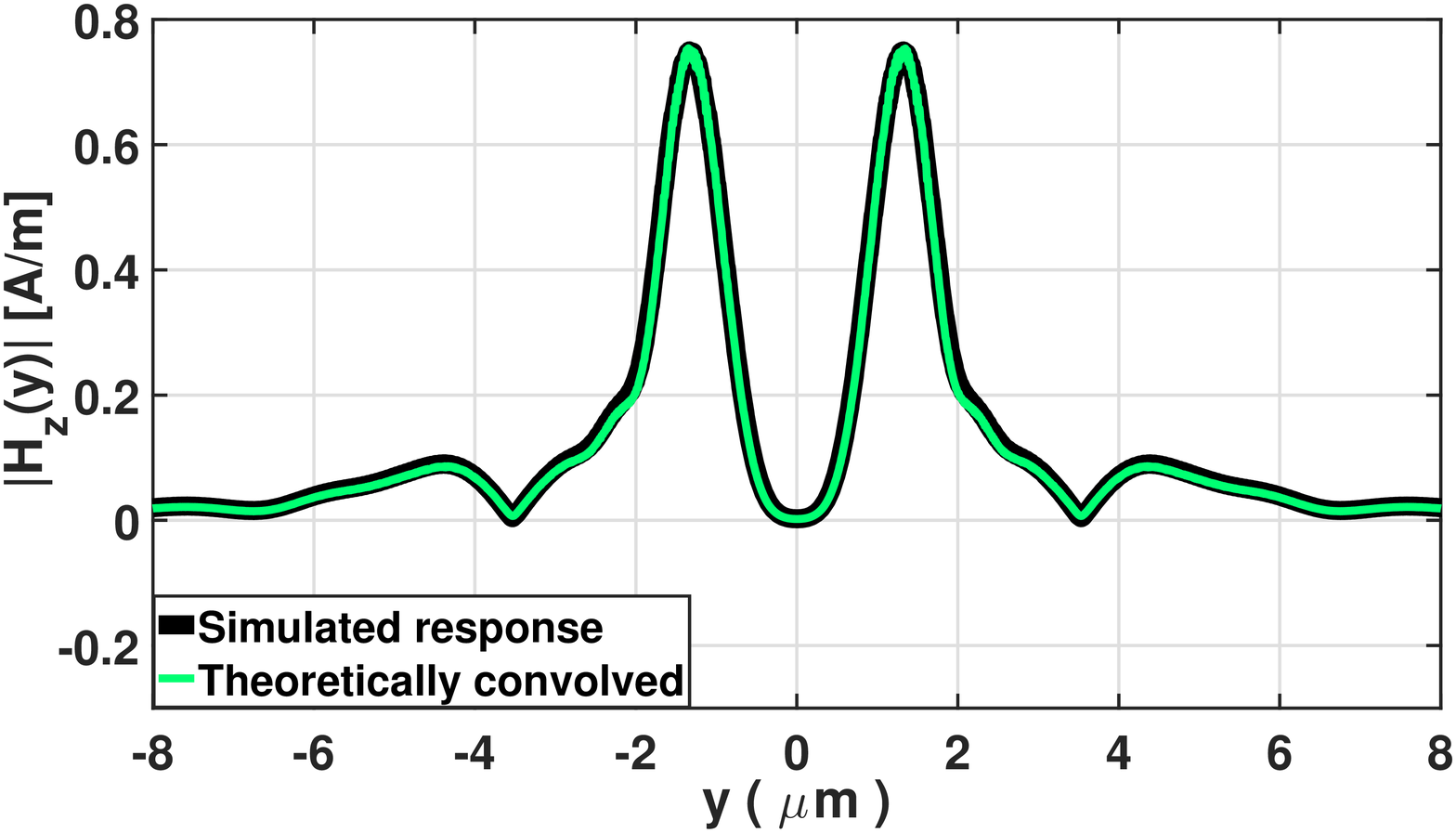}} \\
     \subfloat[\label{subfig-3:Subfig_Convolution_imag}]{\includegraphics[height=0.25\textwidth]{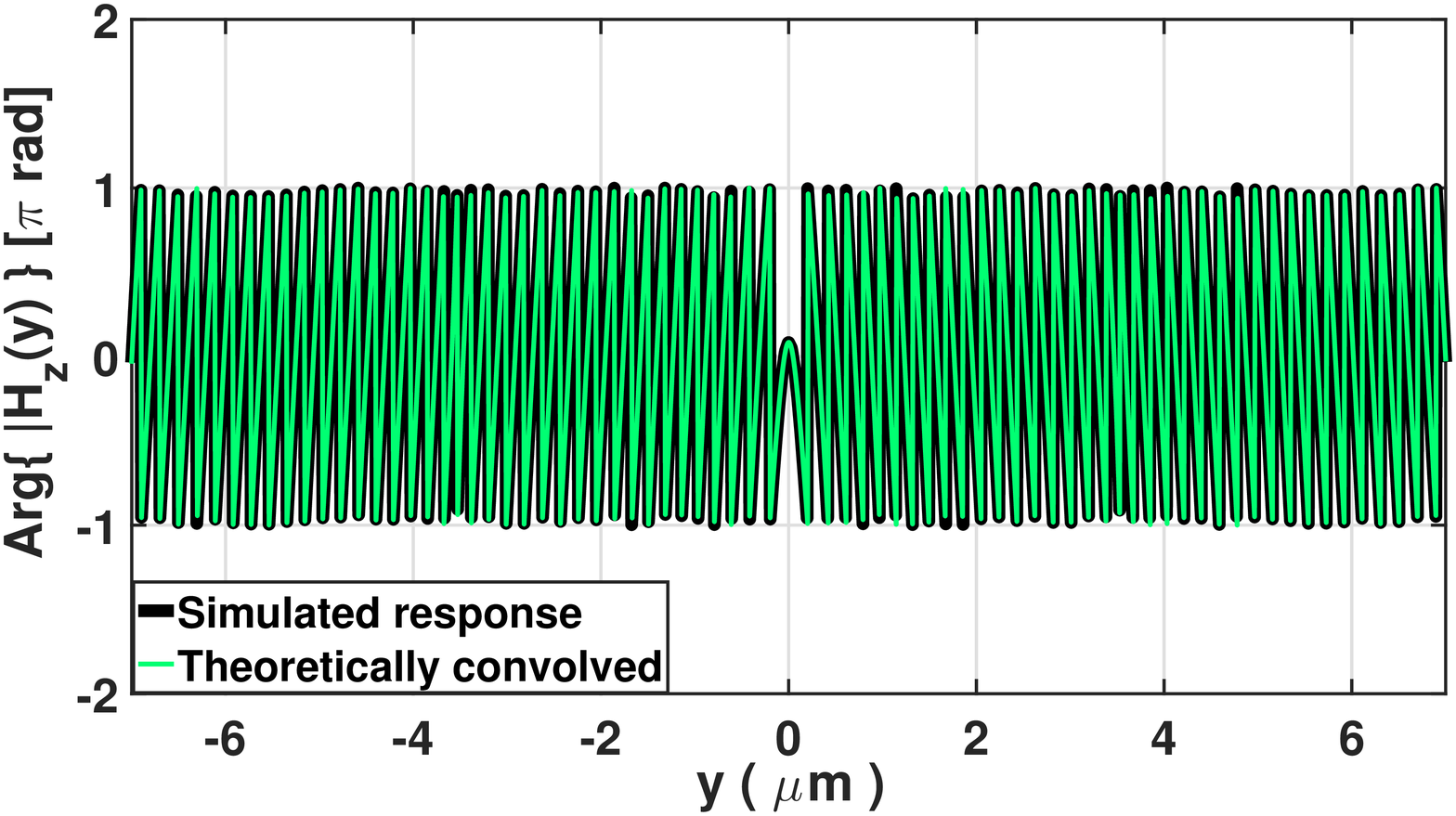}} \\
     \caption{The shift invariance of the multi-layered HMM structure. (a) An arbitrary TM polarized real magnetic field is applied to the geometry. (b) The amplitude  and (c) the phase for the simulated response of the system, shown by the black lines are compared with those from the theoretical convolution shown by the green lines.}
     \label{fig:Fig_Convolution}
\end{figure}

In order to verify the validity of shift invariance, hence the convolution property, we applied an arbitrary TM polarized magnetic field distribution on the input plane. The excitation field is chosen to be purely real and is plotted in figure \ref{subfig-1:Subfig_Arbitrary_field}.  The corresponding response of the system is determined with COMSOL and is extracted as a complex magnetic field distribution from the output plane. The simulated amplitude and phase of the output magnetic field distribution are shown by the black lines in figures \ref{subfig-2:Subfig_Convolution_real} and \ref{subfig-3:Subfig_Convolution_imag}, respectively. If the convolution is satisfied, this simulated response should be equal to the convolution of the input field shown in figure \ref{subfig-1:Subfig_Arbitrary_field} with the PSF of the system shown in figure \ref{fig:Fig_PSF}. The expected response of the system from the convolution is also calculated theoretically and the amplitude and phase of the output magnetic field are shown by the green lines in figures \ref{subfig-2:Subfig_Convolution_real} and \ref{subfig-3:Subfig_Convolution_imag}, respectively. When we compare the simulated response with the convolution result, we can see that there is a very high degree of overlap. This indicates that our assumption of the shift invariance of the multi-layered structure is indeed valid.

\begin{figure}[h]
     \subfloat[\label{subfig-1:Subfig_TF_Amplitude_quartz}]{\includegraphics[height=0.25\textwidth]{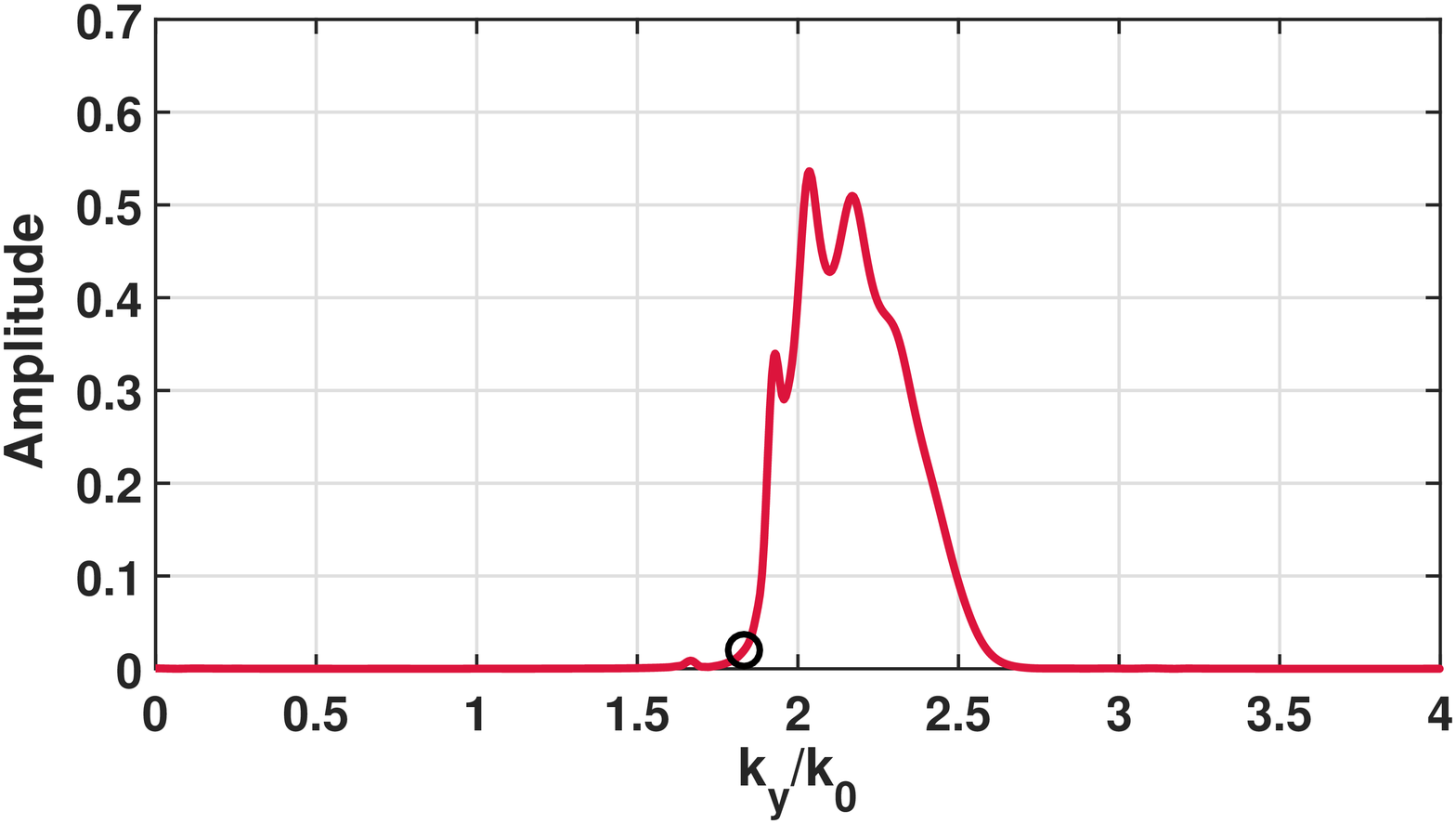}}  \\
     \subfloat[\label{subfig-2:Subfig_TF_Phase_quartz}]{\includegraphics[height=0.25\textwidth]{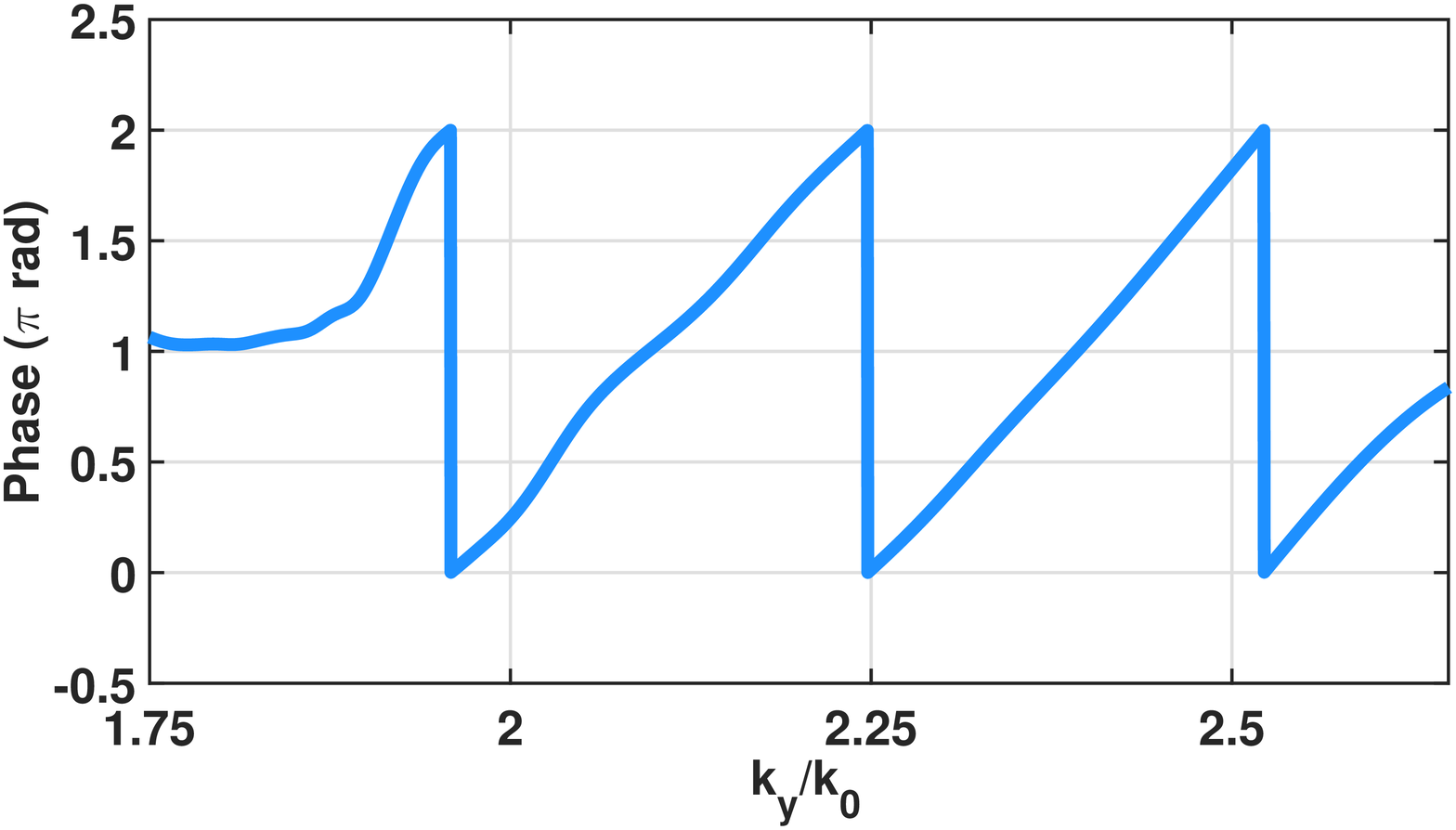}} \\
     \caption{The amplitude and phase of the complex transfer function is shown in (a) and (b), respectively. The circle in (a) corresponds to the cutoff transversal wavevector component calculated from Eq. \ref{eq:Cutoff_frequency}. We see that the wave transmission sharply drops off in this region. Also, the peaks in the transmission spectrum correspond to the eigenmodes of the layered structure. The phase plot shows only the region of high transmission. Outside this region, the transmission is approximately five orders of magnitude less, hence the calculated phase is not reliable.}
     \label{fig:Fig_Complex_transfer_function}
\end{figure}

\begin{figure}[htbp]
\centering
\includegraphics[height=0.25\textwidth]{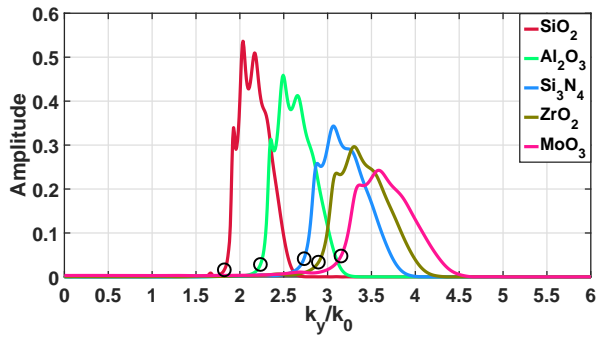}
\caption{The tunable nature of the near-field spatial filter with different dielectrics. The cutoff wavevector components corresponding to each $\epsilon _d$ are calculated from Eq. \ref{eq:Cutoff_frequency} and circled in the respective transmission plots emphasizing the validity of the effective medium for the system.}
\label{fig:Fig_Tunability}
\end{figure}

Having validated the convolution property for the given structure, we move on to verify if the effective medium theory correctly describes the HMM and its spatial filtering behavior. The underlying physics behind the transmission of electromagnetic waves through such a multilayered structure is best explained by the eigenmode approach \cite{zhang2011eigen}. In a layered metal-dielectric system with $N$ metallic layers, there are $N+1$ eigenmodes \cite{li2016hyperbolic}, each having a unique transversal wavevector component. When a radiation field is incident on the medium, only those transversal wavevector components matching the transversal wavevector components of these eigenmodes can resonantly couple into the system. This results in multiple peaks in the transfer function of the multilayered structure \cite{li2017limits}.

We start with the plots of the amplitude and phase of the complex transfer function of the system which are shown in figures \ref{subfig-1:Subfig_TF_Amplitude_quartz} and \ref{subfig-2:Subfig_TF_Phase_quartz}, respectively. Four transmission peaks are clearly visible in the spectrum. Note that only the portion of the phase within the transmission band is shown. The poor transmission on either side means that the phase carries no useful information and is riddled with errors. We note that the phase varies continuously within the transmission band. The apparent discontinuities have a phase change of $2\pi$ which indicates that it is actually continuous. The cutoff wavevector component for the $Al-SiO_2$ multilayered structure can be calculated from Eq. \ref{eq:Cutoff_frequency} and is equal to $k_y = 1.8807k_o$. The corresponding point is circled in the amplitude plot of figure \ref{subfig-1:Subfig_TF_Amplitude_quartz}.  We see that in this region, the transmission drops off rapidly which is consistent with the prediction of the effective medium approximation. Figure \ref{fig:Fig_Complex_transfer_function} clearly shows the spatial filtering property of the HMM around $k_y = 2.2k_0$. In figure \ref{fig:Fig_Tunability} we have used different dielectric materials. The material parameters for these dielectrics \cite{Gao:12,Kelly:72,Luke:15,Wood:82,PhysRevB.88.115141} and their corresponding low cut-off wavevector components are shown in table \ref{tab:Dielectrics}. These cut-offs are indicated by circles in the corresponding transmission plots in figure \ref{fig:Fig_Tunability}. We see that the predictions of the effective medium approximation are still valid since the amplitude transmission drops off sharply below the low cut-offs. Additionally, figure \ref{fig:Fig_Tunability} shows the tunable nature of our proposed spatial filter based on HMMs.

\begin{table}
\centering
\caption{\bf Dielectrics, their permittivities at $\lambda = 365 \ nm$, and low cut-off wavevector components}
\begin{tabular}{c c c}
\hline
\hline
                     &  Relative              & Lower cut-off              \\
Dielectric           &  permittivity          & wavevector component       \\
                     &                        &  $(k_0)$                   \\
\hline
$SiO_2$              & $2.2147$               & $1.88071$ \\
$Al_2 O_3$           & $3.18587$              & $2.28849$ \\
$Si_3 N_4$           & $4.05373$              & $2.77659$ \\
$ZrO_2$              & $5.06205$              & $2.96987$ \\
$MoO_3$              & $6.031 - 1.1908i$      & $3.29314$ \\
\hline
\hline
\end{tabular}
\label{tab:Dielectrics}
\end{table}

\begin{table}
\centering
\caption{\bf Relative permittivity of plasmonic metals at $\lambda = 365 \ nm$.}
\begin{tabular}{c c c}
\hline
\hline
Metal                &  Relative       \\
                     &   permittivity  \\
\hline
Ag   & $-1.8752 -0.59470i$ \\
Al   & $-18.179 -3.2075i$  \\
Pt   & $-4.2933 -8.5848i$  \\
Ta   & $-8.8170 - 9.0576i$ \\
\hline
\hline
\end{tabular}
\label{tab:Plasmonic_metals}
\end{table}

At this point it is important to note that even though the isofrequency contour of Eq. \ref{eq:Dispersion_extraordinary} (see figure \ref{fig:Fig_Hyperbolic}) predicts that an infinite number of transversal wavevector components are allowed in the system, there is however an upper limit. This limit is set by the validity of the effective medium theory which attempts to homogenize the layered system. The approximation ceases to be valid whenever the wavelength corresponding to the longitudinal wavevector component $k_x$ approaches the periodicity of the the layered structure which in our case is $45 \ nm$.

\begin{figure}[htbp]
\centering
\includegraphics[height=0.25\textwidth,width=\linewidth]{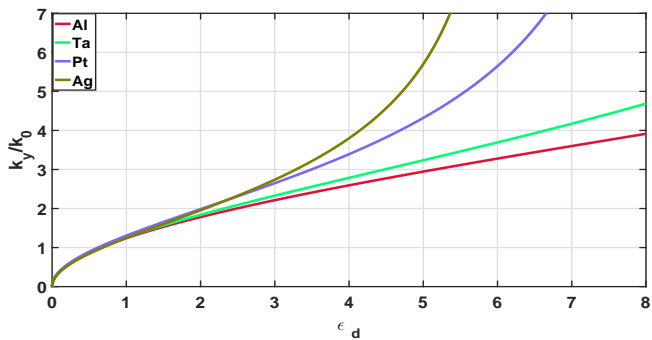}
\caption{Plots of the low cut-off transversal wavevector components versus the dielectric relative permittivity $\epsilon _d$ for different plasmonic metals which give hyperbolic dispersion at $\lambda =365 \ nm$ taking the filling fraction $\eta = 0.5$.}
\label{fig:Fig_Cutoff_compare}
\end{figure}

The general guidelines for designing a near-field band-pass spatial filter can be determined from Eq. \ref{eq:Cutoff_frequency}. Also, figure \ref{fig:Fig_Cutoff_compare} shows the low cut-off wavevector component $k_y^c$ from Eq. \ref{eq:Cutoff_frequency} plotted as a function of the dielectric relative permittivity $\epsilon _d$ for different plasmonic metals at $\lambda = 365 \ nm$. $\eta=0.5$ is kept in all the plots. The permittivities of silver and platinum are calculated from the Drude-Lorentz model in Eq. \ref{eq:Drude_Lorentz} with the data given in \cite{rakic1998optical} whereas the relative permittivity of tantalum is taken from the reflection electron energy-loss spectroscopy data in \cite{werner2009optical}. The relative permittivity data is summarized in table \ref{tab:Plasmonic_metals}. Figure \ref{fig:Fig_Cutoff_compare} can be used to estimate the relative permittivity of the required dielectric material for different plasmonic metals. The slope of the plots is the measure of the sensitivity of the tunable nature of the spatial filter. The sensitivity depends on the selection of both plasmonic metal and the ratio $\eta$. Note that although silver has the highest sensitivity in figure \ref{fig:Fig_Cutoff_compare}, it has a limit beyond which the filter cannot be tuned with dielectrics. This is due to the loss of the hyperbolic nature of the layered structure above a certain value of the relative permittivity $\epsilon _d$. In contrast, other metals allow tunability for a broader range of transversal wavevector components at the expense of higher loss and stringent dielectric permittivity requirements. While loss can be mitigated by controlling the filling fraction or the number of unit cells to some extent, the requirement of large permittivity imposes a limitation on the tunability especially at small optical wavelengths.
Note, however, that the proposed spatial filter can be scaled to different wavelengths as long as the layered system exhibits hyperbolic dispersion. This is because the normalized cutoff wavevector component in Eq. \ref{eq:Cutoff_frequency} is not explicitly dependent on the wavelength and varies according to the material parameters and filling fraction.

Finally, it would be appropriate to conclude this section by discussing the integration of this spatial filter into a near-field imaging system to generate the auxiliary source in \cite{Ghoshroy:17}. The auxiliary source is constructed from the convolution of the object, hence inherits partial information about the object.  At first glance it might seem that this compromises the entire scheme for one is tempted to ask ``how is it possible to incorporate features of the object into the auxiliary without any prior knowledge about the object?" The answer to this question is surprisingly simple. One simply investigates the Fourier spectrum of the raw image and determines the spatial frequency region where the noise has degraded the signal. One then uses the proposed spatial filter tuned to a small band of high-frequency Fourier components in the noise dominated spectral region. When this filter is placed on the object and illuminated with sufficiently strong intensity of coherent light an immediate consequence is the selective amplification of the signal within the ``pass-band" (see figure \ref{fig:Fig_Tunability}) of the filter \cite{Ghoshroy:17}. The process can be repeated iteratively by tuning this pass-band  to another noise dominated region in the Fourier spectrum of the image with no prior information about the object. The details of the entire iterative process can be found in \cite{Ghoshroy:17}.

\section{Conclusion}

In conclusion, we have proposed a near-field spatial filter for the active implementation \cite{Ghoshroy:17} of the recently introduced $\Pi$ loss compensation scheme \cite{sadatgol2015plasmon}. The ``tunability" and ``selective amplification" characteristics of the auxiliary source in \cite{Ghoshroy:17} can be realized with layered metal-dielectric systems with hyperbolic dispersion. We have demonstrated that the convolution, which is vital for the construction of the auxiliary source, can be achieved in the layered system. This allows such layered systems to be integrated with the near-field superlenses (e.g., silver \cite{fang2005sub} and SiC \cite{taubner2006near} lenses), so that the complete imaging system can be described with a modified transfer function. This work paves the way to a robust loss compensation scheme for enhanced near-field superlensing with ultra-high resolution. The future work will focus on the demonstration of how the proposed spatial filter, when integrated with near-field superlenses can be used to reconstruct images deep beyond the diffraction limit without any prior knowledge of the object. Interestingly, a spatial filter of this form may also have potential applications in “edge-detection” as proposed for acoustics in a recent work \cite{moleron2015acoustic}, where an acoustic metamaterial is used to transmit the high-spatial evanescent modes while suppressing the propagating modes. The authors allude to the lack of similar systems in optics. It is important to note that this is exactly what our proposed near-field spatial filter does in the optical domain.

\section*{Acknowledgements}

This work was supported by Office of Naval Research (award N00014-15-1-2684). Superior, a high-performance computing infrastructure at Michigan Technological University, was used in obtaining the results presented in this publication.

\end{document}